\documentclass[%
 aip,
 jcp,%
 amsmath,amssymb,
 reprint,%
]{revtex4-1}

\usepackage{graphicx}% Include figure files
\usepackage{dcolumn}% Align table columns on decimal point
\usepackage{bm}% bold math

\newcommand{\st}{\bm{\sigma} ( \mathbf{r} ) }

\begin{document}

%\preprint{AIP/123-QED}

\title[]{The Lorentz sphere visualised}% Force line breaks with \\

\author{S. Sturniolo}
\email{simone.sturniolo@stfc.ac.uk}
\affiliation{Scientific Computing Department, UKRI}
\author{J.R. Yates}
\email{jonathan.yates@materials.ox.ac.uk}
\affiliation{Department of Materials, University of Oxford, Parks Road, Oxford, OX1 3PH, United Kingdom}

\date{\today}% It is always \today, today,

\begin{abstract}
From the inception of Nuclear Magnetic Resonance as a spectroscopic technique, the local origin of chemical shifts has been a topic of discussion. A useful concept employed to describe it has been that of the ``Lorentz sphere'', the approximately spherical volume surrounding a given nucleus in which the electronic currents contribute significantly to the chemical shift, whereas the outside can be considered as an uniformly magnetised ``bulk''. \newline
In this paper we use the output of the plane wave DFT code CASTEP to get a quantitative estimate of the Lorentz sphere in periodic systems. We outline a mathematical description of a radial buildup function for the magnetic shielding starting from the electronic currents and the simple assumption of periodicity. We provide an approximate upper bound for the Lorentz sphere's size in any crystal, then compute buildup functions for a number of sites in two molecular crystals, showing how various chemical features such as hydrogen bonds influence to convergence to the final shielding value.
\end{abstract}

%Use showkeys class option if keyword
\maketitle

\section{\label{sec:intro}Introduction}

Since the early days of NMR it has been vital to understand the way in which chemical structure gives rise to chemical shifts \cite{zimm_1957}. Physical considerations lead one to separate two main contributions: one originating from the bulk of the sample, an average effect of macroscopic magnetisation and sample shape, and another coming from local electronic currents, whose structure on the short range causes the observed differences between sites in the same sample. In molecules tumbling around in a solvent in the liquid state, there is a natural separation between intra- and intermolecular contributions for these local effects, due to motional averaging of the latter. In the solid state, however, this distinction is more blurred. The range within which these local contributions are relevant is referred to as the ``Lorentz sphere'' in literature \cite{zimm_1957, durr_2005, ulrich_2003}, but such considerations typically do not go beyond a qualitative level. It would be very useful for the purpose of NMR crystallography to have a better, quantitative understanding of the Lorentz sphere, as it would allow us to better understand how much and in what ways each spectral line is sensitive to the overall structure.\newline
The authors have already suggested one such approach \cite{zilka_2017} making use of electronic currents computed by Density Functional Theory, in which the volumes of space contributing the most to the chemical shift at a given site were highlighted. This method was similar to the approach of plotting current density maps that has been used in the past \cite{jameson_1979, jameson_1980, cuesta_2009}. In this work however a different approach, specifically suited to crystalline solids, is proposed. The properties of the current in a periodic crystal are exploited to compute a radial buildup function. As we expand the volume of the crystal whose contribution we account for in a sphere centred on the site, we see how shielding tensors converge to their final value. The analytic form of this function gives an upper bound for the size of the Lorentz sphere that depends only on the lattice parameters of the crystal itself. Finally, Density Functional Theory (DFT) and the Gauge-Including Projector Augmented Wave (GIPAW) approach\cite{pickard_2001,yates_2007,Bonhomme2012} are used to compute the currents induced by valence electrons in two well known molecular crystals, ethanol and aspirin, and buildup functions are created for isotropic shielding, anisotropy and asymmetry at various sites. The shape and speed of convergence of these curves highlights the way each site interacts with its chemical environment, confirming physical intuition and providing a strikingly simple visual representation of the Lorentz sphere.

\section{\label{sec:theory}Theory}

Applying a static magnetic field to a diamagnetic sample will induce microscopic electronic currents. These currents give rise to local magnetic fields which are experienced by atomic spins and shift slightly their resonance frequencies, thus generating the NMR spectrum. If we assume that the current density, $\mathbf{J}(\mathbf{r}')$ is known, then the magnetic shielding tensor at a given position, $\st$, describing the relationship between external applied field $\mathbf{B}_{ext}$ and local field $\mathbf{B}_{in}$, can be found simply by applying the Biot-Savart law \cite{yates_2007}:

\begin{equation}\label{theory_bsavart}
    \mathbf{B}_{in}(\mathbf{r}) = -\st\mathbf{B}_{ext} = \frac{1}{c}\int d^3 r' 
    \mathbf{J}(\mathbf{r}')\times\frac{\mathbf{r}-\mathbf{r}'}{|\mathbf{r}-\mathbf{r}'|^3}
\end{equation}

where the integral is carried out over the entirety of space. In the case of an infinite periodic system, however, we can exploit the properties of the current density to write it as a Fourier series over the reciprocal lattice vectors $\mathbf{G}$:

\begin{equation}\label{theory_jfourier}
    \mathbf{B}_{in}(\mathbf{r}) = \frac{1}{c}\int d^3 \rho 
    \frac{\bm{\rho}}{\rho^3} \times \sum_\mathbf{G}
    \mathbf{J}(\mathbf{G})\exp(i\mathbf{G}\bm{\rho})\exp(i\mathbf{G}\mathbf{r})
\end{equation}

where a change of coordinate $\bm{\rho} = \mathbf{r}'-\mathbf{r}$ has been made for convenience.\newline
Passing to polar coordinates, and using the plane wave expansion \cite{jackson_1998}:

\begin{align}\label{theory_intpolar}
    \mathbf{B}_{in}(\mathbf{r}) &= \frac{1}{c}\int \rho^2 d \rho d\Omega 
    \frac{\hat{\bm{\rho}}}{\rho^2} \times \\
    & \sum_\mathbf{G}
    4\pi\mathbf{J}(\mathbf{G})\exp(i\mathbf{G}\mathbf{r})\sum_{l,m}i^l j_l(G\rho)
    Y_{l,m}(\hat{\mathbf{G}})Y_{l,m}^*(\hat{\bm{\rho}}) \nonumber
\end{align}

where the $Y_{l,m}$ and $j_l$ are respectively the spherical harmonics and the spherical Bessel function with angular momentum numbers $l, m$.\newline
Since the $\hat{\bm{\rho}}$ versor can be entirely expressed as a function of real spherical harmonics with $l=1$:

\begin{equation}\label{theory_rho_spharm}
    \hat{\bm{\rho}} = \left[Y_{1,-1}-Y_{1,1}, i(Y_{1,-1}+Y_{1,1}), \sqrt{2}Y_{1,0}\right]
\end{equation}

(where dependency of the Ys on $(\hat{\bm{\rho}})$ is left implicit), when integrating over the solid angle $\Omega$ the sum over $l,m$ can be removed as all other terms disappear, and these are projected and replaced with the corresponding spherical harmonics dependent on $\hat{\mathbf{G}}$. In the end we have

\begin{align}\label{theory_intradial}
    \mathbf{B}_{in}(\mathbf{r}) &= \frac{1}{c} 
    \sum_\mathbf{G}
    4\pi i \hat{\mathbf{G}}\times\mathbf{J}(\mathbf{G}) \exp(i\mathbf{G}\mathbf{r}) \\
    &    \int d \rho  j_1(G\rho)\nonumber.
\end{align}

Exploiting the properties of the Bessel functions:

\begin{equation}\label{theory_bess_prop}
    j_1(x) = -\frac{\partial}{\partial x} \frac{\sin(x)}{x}
\end{equation}

we find

\begin{equation}\label{theory_bess_int}
    \int_0^R d \rho  j_1(G\rho) = -\frac{1}{G}\left[\frac{\sin(GR)}{GR}-1\right]
\end{equation}

leading to

\begin{align}\label{theory_buildup}
    \mathbf{B}_{in}(\mathbf{r}, R) &= \frac{1}{c} 
    \sum_\mathbf{G}
    4\pi i \hat{\mathbf{G}}\times\mathbf{J}(\mathbf{G}) \exp(i\mathbf{G}\mathbf{r}) \\
    &   \frac{1}{G}\left[1-\frac{\sin(GR)}{GR}\right] \nonumber.
\end{align}

In the limit of $R\rightarrow\infty$ this reduces to the known result \cite{yates_2007}

\begin{equation}\label{theory_inflimit}
\mathbf{B}_{in}(\mathbf{r}) = \frac{1}{c} 
    \sum_\mathbf{G}
    4\pi i \frac{\hat{\mathbf{G}}}{G}\times\mathbf{J}(\mathbf{G}) \exp(i\mathbf{G}\mathbf{r})
\end{equation}

which can be obtained from  Eq. \ref{theory_bsavart} also simply by applying the convolution theorem to the individual Fourier transforms of the two factors of the integral.\newline
Equation \ref{theory_buildup} gives us a way to compute a buildup function for the magnetic shielding tensor at a given site, assuming that the electronic current density is known. The only expensive operation is computing the current density itself; given that, Eq. \ref{theory_buildup} is trivial to compute at any value of $R$ in just a few seconds. Full buildup functions with good resolution can be obtained in minutes for average sized systems. Even without considering the details of the current density for a specific system, though, we can make some useful observations just from the form of the expression. The sum over all $\mathbf{G}$ vectors appearing in  Eq. \ref{theory_jfourier} to \ref{theory_inflimit} is a sum over all the points of the reciprocal lattice, except for $\mathbf{G} = 0$, since we assume that there is no net current in the crystal. This means that the convergence of the sum in Eq. \ref{theory_buildup} will be often dominated by the shortest $\mathbf{G}$ in the lattice, for two reasons. The first is that the $\frac{1}{G}$ factor significantly enhances its relative magnitude. The second is that the Bessel function $sin(x)/x$ converges quickly to zero as $x$ goes to infinity; for $x \sim 6$ it is already within a $20\%$ error of its final value. Therefore, the higher $G$, the faster the convergence of a specific term. One can consider the simple case of a cubic lattice with parameter $a$ and see that the argument will be $x=2\pi R/a$ and therefore will be already within the $20\%$ error for $R\sim a$. In the more general case, Eq. \ref{theory_buildup} suggests that the size of the Lorentz sphere in a periodic system will tend to be on the scale of the inverse of the length of the shortest reciprocal lattice vector. This is by no means a hard rule, as in special cases there could be constructive interference effects that cause the convergence to be slower. The exact size of the sphere will always depend on the level of precision for which we want to account and the current density of the system of interest.

\section{\label{sec:computation}Computation of buildup functions}

During the last two decades, routine computation of magnetic shieldings by use of Density Functional Theory (DFT) software has been made possible by the development of the Gauge-Including Projector Augmented Wave (GIPAW) method \cite{pickard_2001, yates_2007}. This method solves the issue of computing shieldings when representing the core electrons of an atom with pseudopotentials. This is a required step to only run the DFT calculation proper to compute the wavefunction of the valence electrons, thus decreasing the computational load significantly. When operating within this approximation, the final computed magnetic field can be split in three parts:

\begin{equation}\label{comp_bfield}
    \mathbf{B}_{tot} = \mathbf{B}_{in} + \mathbf{B}_{susc} = \mathbf{B}_{loc} + \mathbf{B}_{lr} + \mathbf{B}_{susc}.
\end{equation}

Here $\mathbf{B}_{susc}$ represents a contribution from the susceptibility of the sample, which is simply a function of its shape, and $\mathbf{B}_{loc}$ and $\mathbf{B}_{lr}$ represent the local and long range contributions respectively. The local contribution here is used to mean an extremely short-ranged contribution due to the core electrons and to part of the valence electons. This is a consequence of the way the GIPAW approximation works, as the wavefunction near the nucleus is pseudised. Therefore, in the upcoming examples we will focus only on the long-range, which is the one that contains all the effects of valence electrons away from the nucleus. This means that the buildup functions should not considered fully physical within the core radius for each element, which are 0.32 \AA{} for H, 0.74 \AA{} for C and 0.58 \AA{} for O.

\subsection{\label{subsec:comp_details}Details of the calculations}

Calculations were carried out using CASTEP version 16.1 \cite{clark_2005}. Structures for solid state ethanol \cite{jonsson_1976} and aspirin (acetylsalicylic acid) \cite{wilson_2002} were retrieved from the Crystallography Open Database. Calculations were carried out using the PBE exchange-correlation functional with Tkatchenko-Scheffler dispersion corrections \cite{mcnellis_2009} and the default ultrasoft pseudopotentials. After convergence checks, the cutoff energy was set to 800 eV for ethanol and 600 eV for aspirin; the k-point grid was set to 3x2x2 for ethanol and 2x3x2 for aspirin. A full geometry optimisation to a tolerance of 0.01 eV/\AA{} was carried out for both systems, and then the valence electron current densities were calculated and written in a binary file using CASTEP's Magres module \cite{pickard_2001, yates_2007} with the keyword \texttt{MAGRES\_WRITE\_RESPONSE = TRUE}.\newline
Shielding buildup functions were constructed using a Python script making use of the NumPy, SciPy\cite{oliphant_2007, millman_2011} ASE \cite{larsen_2017} and Soprano \cite{soprano} libraries for computing and parsing of structure files. The code has been made available on GitHub, currently still as a prototype, with the name \texttt{pynics} \cite{pynics}.\newline
Full output structures, the coordinates of the atoms for which buildup functions were calculated, and all other results are provided as supplementary information.

\subsection{\label{subsec:comp_conventions}Conventions and notation}

We clarify here some of the notation used throughout this paper. First, while most NMR literature references chemical shifts, in this paper we will compute and plot magnetic shieldings. The difference between these two quantities is simply a constant. Magnetic shielding tensors express the absolute magnetic response to an external field experienced at a site; chemical shifts are the same response referenced to that of another, known sample. Therefore, we will focus on magnetic shieldings, which are the most natural output of a DFT calculation.\newline
The magnetic shielding is a rank 2 tensor with nine independent components, though for most NMR experiments only the symmetric part is of interest, reducing the independent components to only six. In order to express them more conveniently, the so-called Haeberlen convention is used \cite{haeberlen1976advances, mehring1983nuclear, spiess1978nmr}. In this convention, the eigenvalues of the shielding tensors are computed. The isotropic shielding is defined as their average, or one third of the trace of the tensor:

\begin{equation}\label{conv_iso}
    \sigma_{iso} = \frac{\sigma_1+\sigma_2+\sigma_3}{3}.
\end{equation}

The eigenvalues are ordered so that

\begin{equation}\label{conv_haeb}
    |\sigma_3-\sigma_{iso}| > |\sigma_1-\sigma_{iso}| > |\sigma_2-\sigma_{iso}|
\end{equation}

and two other quantities, anisotropy and asymmetry, are defined respectively as:

\begin{equation}\label{conv_aniso}
    \sigma_{aniso} = \sigma_3 - \frac{\sigma_1+\sigma_2}{2}
\end{equation}

\begin{equation}\label{conv_asymm}
    \sigma_{asymm} = \frac{\sigma_2-\sigma_1}{\sigma_{aniso}}.
\end{equation}

\begin{figure}[ht]
    \centering
    \includegraphics{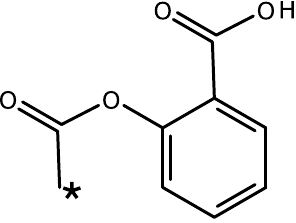}
    \caption{Aspirin molecular structure. The $\mathrm{CH}_3$ group is labelled with a star.}
    \label{fig:aspirin_sites}
\end{figure}

Hydrogen atoms will be referred to by the site they are bonded to. When discussing ethanol, this means they will be referred to as $\mathrm{CH}_3$, $\mathrm{CH}_2$ and $\mathrm{OH}$ hydrogens respectively. Despite being chemically equivalent, these sites are not crystallographically equivalent, and have different shieldings; however, they retain enough similarity that treating them in groups still makes sense.\newline
For aspirin, the situation is more complex. In figure \ref{fig:aspirin_sites} a molecule of aspirin is represented; since the only site that was analysed in this case is the $\mathrm{CH}_3$ group, it is labelled appropriately.

\subsection{\label{subsec:comp_results}Results}

Shielding buildup functions were computed from the DFT valence electron current densities for all atomic sites in one molecule of ethanol and aspirin inside the given crystal structures. Table \ref{tab:res_lattice} shows the final lattice parameters for the optimised structures as well as the inverse of the shortest reciprocal lattice vector, which should give us a scale for convergence of the shielding.

\begin{table}[ht]
    \centering
    \begin{tabular}{|c|c|c|c|c|}
    \hline
        Molecule & a (\AA) & b (\AA) & c (\AA) & $1/\min(|G|)$ (\AA) \\
    \hline
        Ethanol & 5.33 & 6.81 & 8.06 & 1.25 \\
    \hline
        Aspirin & 11.32 & 6.54 & 11.38 & 1.8 \\
    \hline
    \end{tabular}
    \caption{Lattice parameters and inverse shortest reciprocal lattice vector for ethanol and aspirin.}
    \label{tab:res_lattice}
\end{table}

\begin{figure}[ht]
    \centering
    \includegraphics[width=0.5\textwidth]{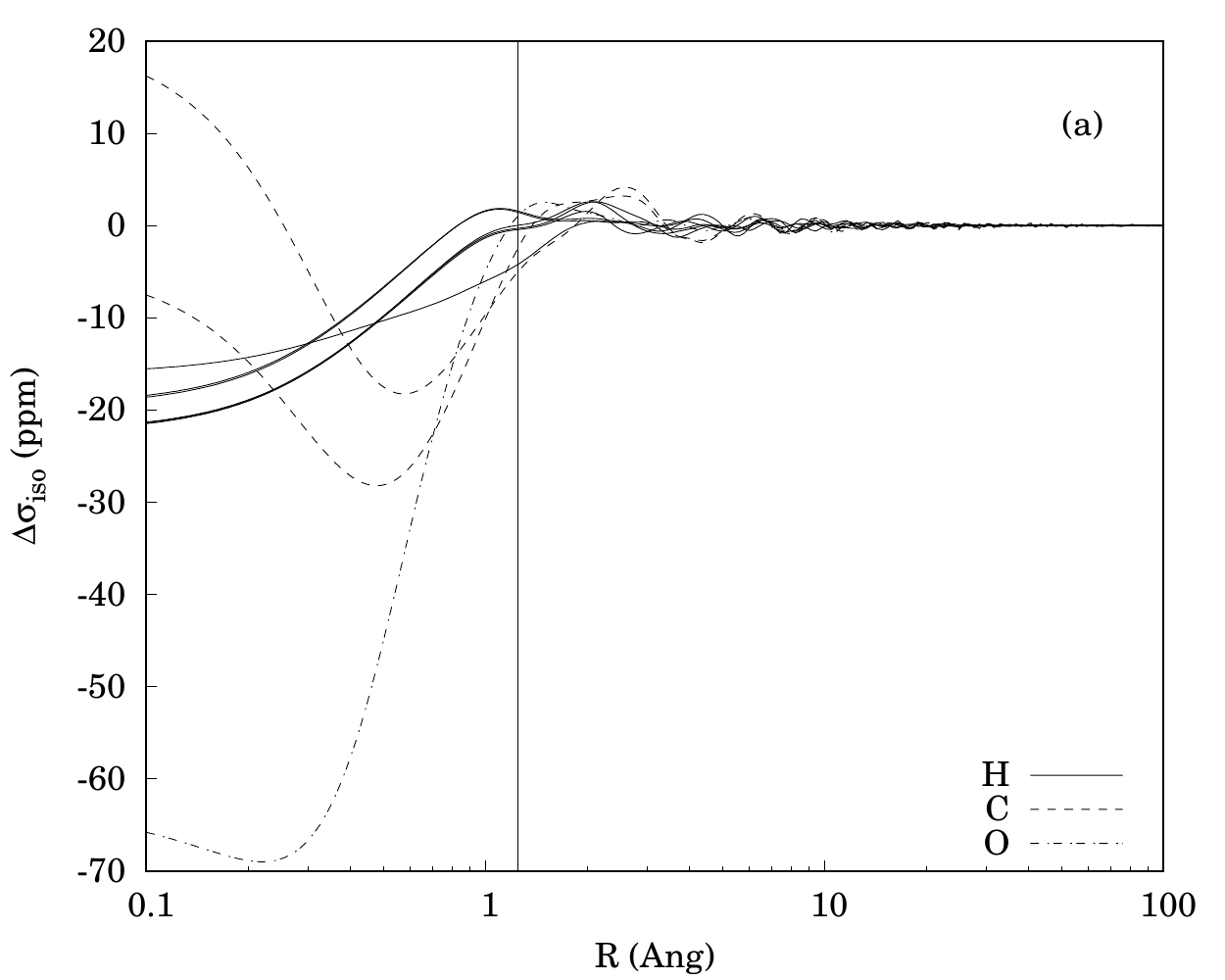}
    \includegraphics[width=0.5\textwidth]{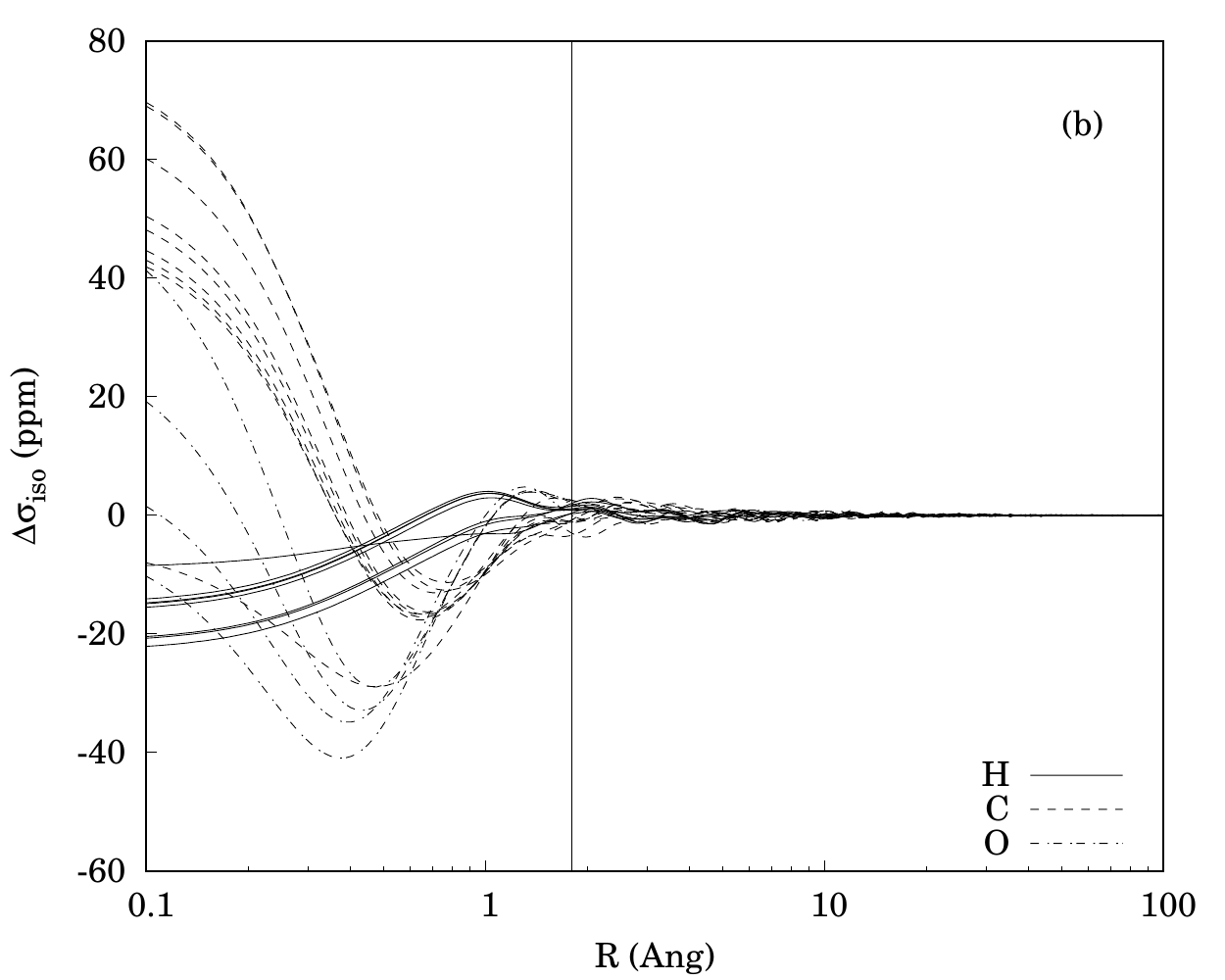}
    \caption{Long range convergence of $\sigma_{iso}(R)-\sigma_{iso}$ for all nuclei in one ethanol (a) and one aspirin (b) molecule, grouped by species, in their respective crystals. The value of $1/\min(|G|)$ for each system is marked with a vertical line}
    \label{fig:aspirin_conv}
\end{figure}

\begin{figure}[ht]
    \centering
    \includegraphics[width=0.5\textwidth]{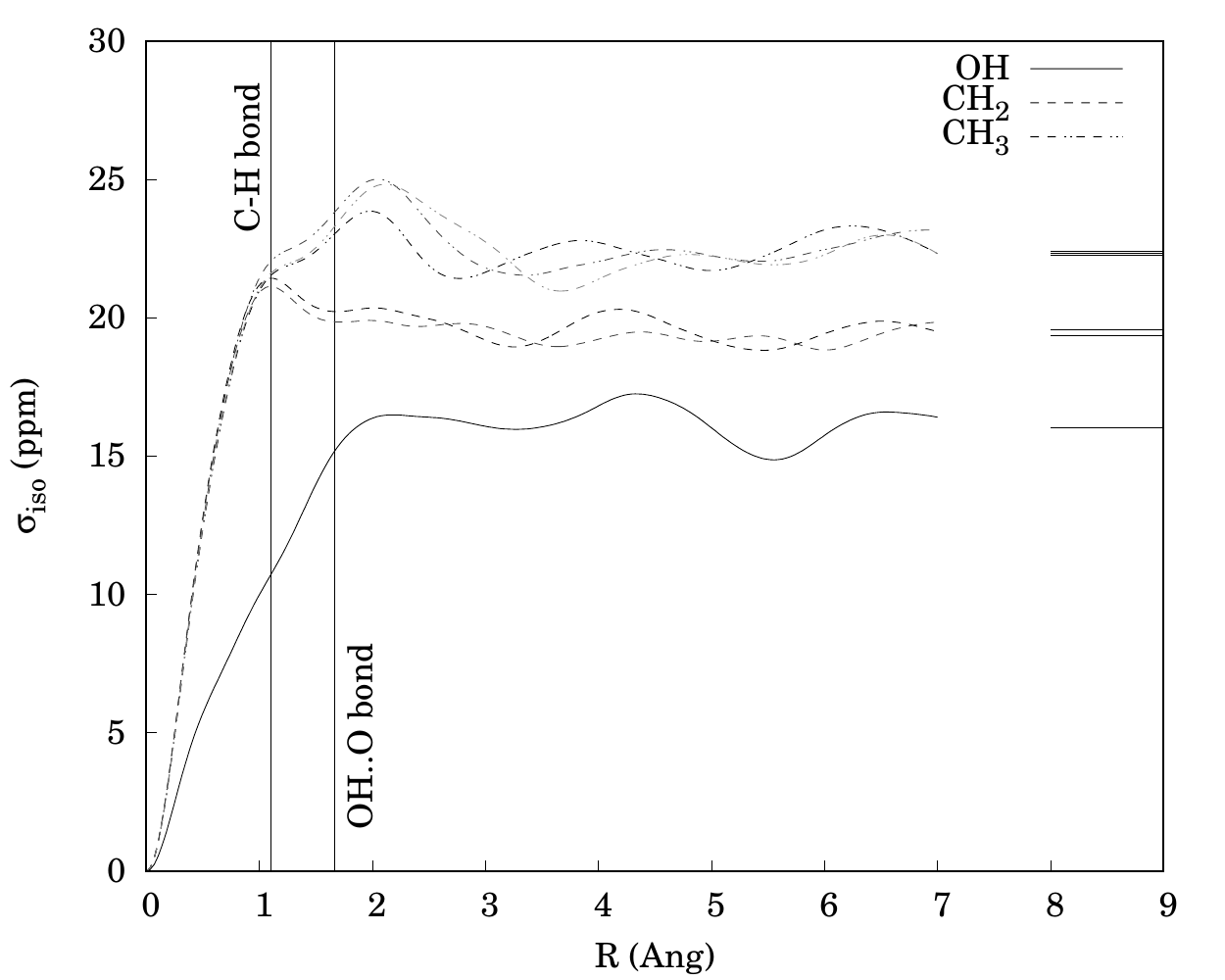}
    \caption{Isotropic shielding buildup functions for hydrogen atoms in ethanol, by chemical group. On the right, the predicted final values are shown as lines. Vertical lines are used to mark the typical length of a C-H covalent bond and of the distance between the hydrogen and the acceptor oxygen in the hydrogen bond.}
    \label{fig:res_eth_protons}
\end{figure}

 In Figure \ref{fig:aspirin_conv}, the buildup functions for all atoms in ethanol and aspirin up to a distance of 100 \AA{}
  are shown, expressed as deviation from the final value, $\Delta\sigma_{iso}(R) = \sigma_{iso}(R)-\sigma_{iso}$. This plot makes the scale of the convergence apparent. For most nuclei, the largest part of their shielding converges to a value close to the final one on a very short radius, comparable with the predicted convergence scale $1/\min(|G|)$. Smaller oscillations continue up to a much larger range, but ultimately, they become imperceptible far before the 100 \AA{} mark. This confirms the initial intuition that the large part of the shielding is justified by local currents. More insights about the origin of various contributions to the shielding can be gained by observing the buildup functions on a smaller scale.\newline
Ethanol gives us a good insight in the way in which inter-molecular interactions affect the size of the Lorentz sphere. In Figure \ref{fig:res_eth_protons} we can see the buildup functions for the shielding isotropy on each of the six protons in an ethanol molecule. The curves are grouped by carbon site that the protons are bonded to. One obvious observation is that five of the curves converge much faster than the last one, the one for the OH proton. This is due to the fact that this proton participates in a hydrogen bond with the oxygen from another molecule. The distance between the proton and the acceptor oxygen is of 1.67 \AA{}, which corresponds roughly to the point where the curve converges to a value very close to its final one. This long distance effect agrees with the observations made in a similar situation in a previous paper by the authors \cite{zilka_2017}. The other five shieldings converge much quicker, on a length comparable to 1.1 \AA{}, which is the length of the C-H covalent bond. It is also interesting to note that these five curves basically overlap on that distance, reflecting the close similarity of the local electronic wavefunction for these five sites, and diverge in distinct groups for the $\mathrm{CH}_2$ and $\mathrm{CH}_3$ protons only afterwards, influenced differently by the rest of the molecule. To the right side of the figure, the final shieldings are marked by short segments, showing how the curves are almost effectively converged to their final values.

\begin{figure}[ht]
    \centering
    \includegraphics[width=0.5\textwidth]{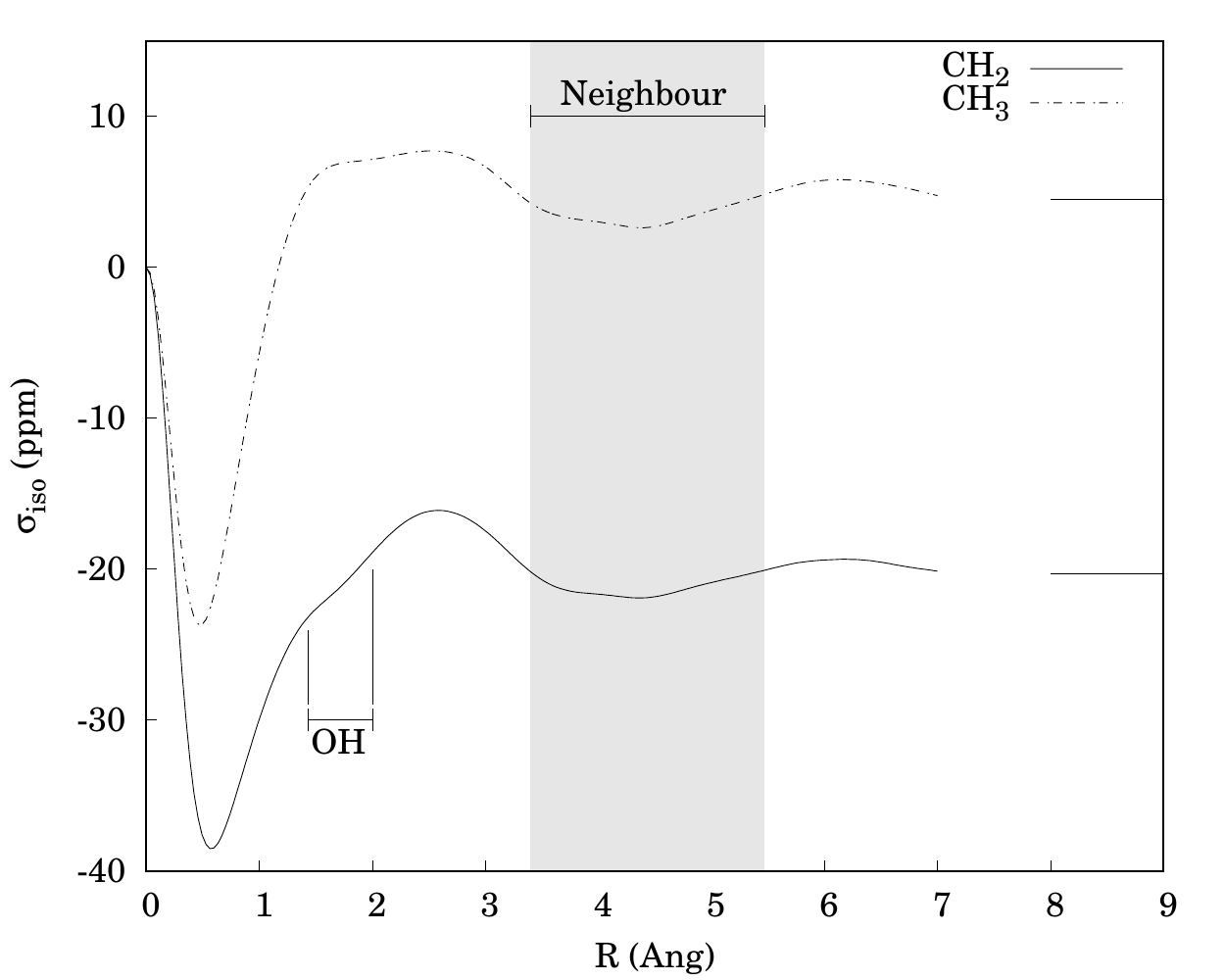}
    \caption{Isotropic shielding buildup functions for carbon atoms in ethanol, by chemical group. On the right, the predicted final values are shown as lines. The range of distances at which the OH group appears with respect to the $\mathrm{CH}_2$ one, and the general range of distances of the next closest neighbouring molecule for both carbon atoms, are highlighted.}
    \label{fig:res_eth_carbons}
\end{figure}

In Figure \ref{fig:res_eth_carbons} we see the buildup functions for the two carbons in ethanol. These curves are very similar in shape, but end up leading to two very distinct final values because of two major differences. The first is that the $\mathrm{CH}_2$ carbon curve descends much quicker and lower on a very short range, within the core radius of the carbon pseudopotential. The second difference is a bump that is visible around the 2 \AA{} mark, in a range highlighted in the figure and corresponding approximately to the range of distances comprising the OH group from the $\mathrm{CH}_2$ carbon's position. This suggests that the origin of this bump is effectively the electronic cloud surrounding that group. The carbon curves also show much better than the proton ones the effect of the rest of the crystal. A feature is visible starting around 3.4 \AA{}. This is approximately the distance from either of the carbon atoms and the closest atom of a different molecule from the one they belong to. The range of distances representing the closest neighbouring molecule is highlighted. This shows very clearly the importance of the direct effect of the crystal on the shielding, which appears to be similar for both sites.

\begin{figure}[ht]
    \centering
    \includegraphics[width=0.5\textwidth]{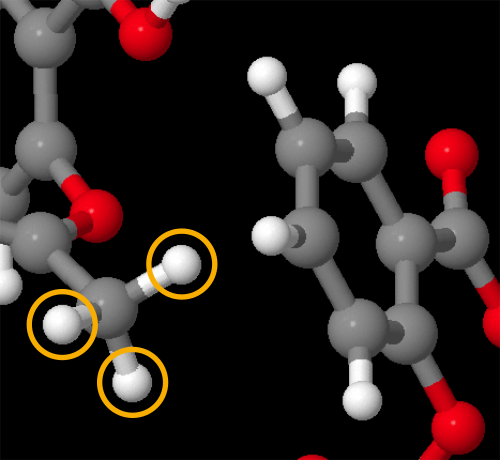}
    \caption{Detail of aspirin crystalline structure with the $\mathrm{CH}_3$ protons highlighted. The neighbouring aromatic ring is visible on the right.}
    \label{fig:res_eth_aspirin_cluster}
\end{figure}

\begin{figure}[ht]
    \centering
    \includegraphics[width=0.5\textwidth]{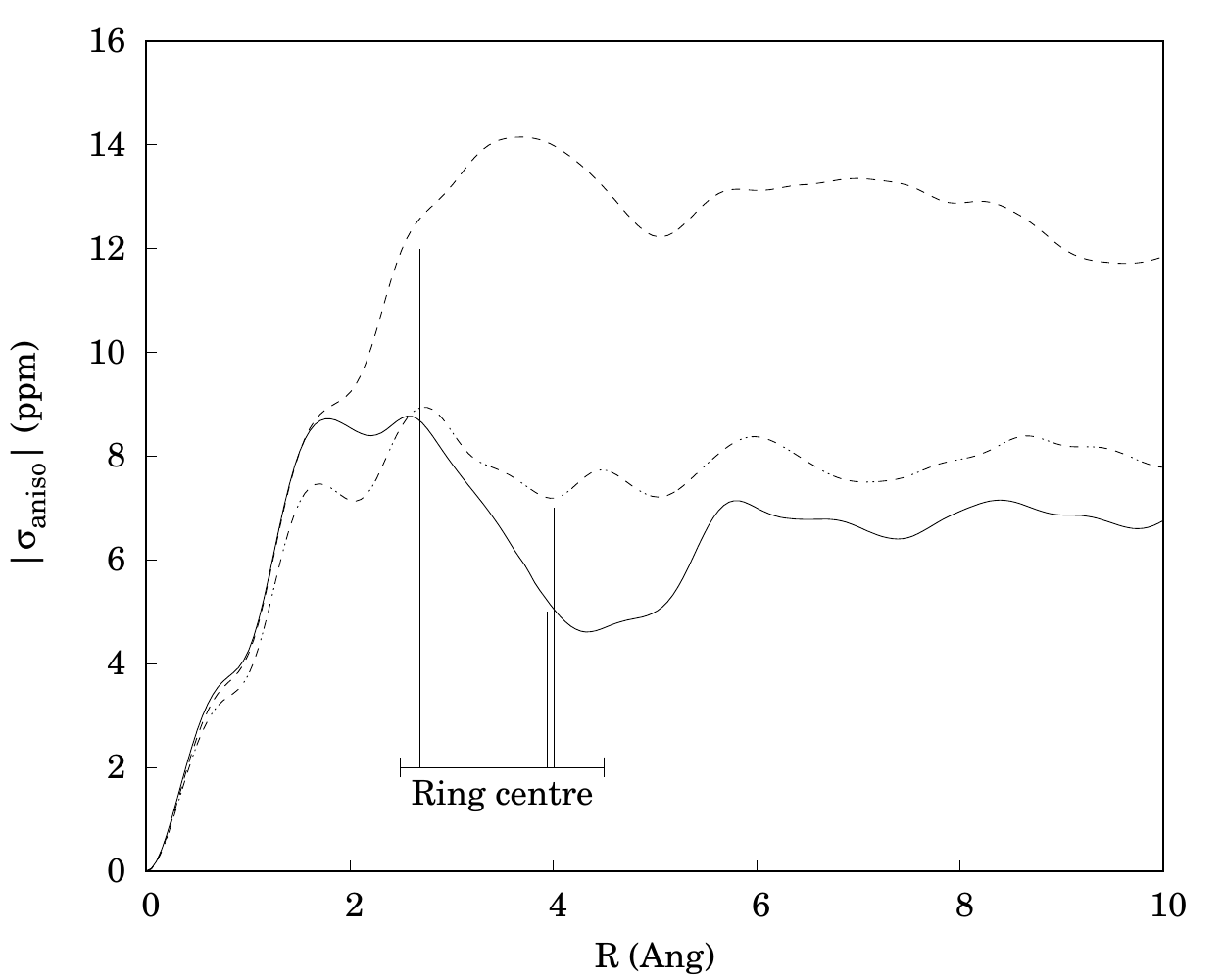}
    \caption{Absolute value of the shielding anisotropy for the $\mathrm{CH}_3$ protons in aspirin. The distances of each proton from the centre of the nearby aromatic ring are marked by lines.}
    \label{fig:res_eth_aspirin_plot}
\end{figure}

Aspirin has been chosen as a test example because of the presence of one aromatic ring in its structure, which is likely to cause strong intermolecular effects. This is most visible in the protons belonging to the $\mathrm{CH}_3$ group. The arrangement of these protons is shown in Figure \ref{fig:res_eth_aspirin_cluster}. All three protons are within a radius of 4 \AA{} from the centre of the nearby molecule's aromatic ring, but one of them is much closer than the other two, and its C-H bond is aligned in a direction almost orthogonal to the plane of the ring. To highlight the effect of the ring currents, which is strongly dependent on direction, instead of plotting the isotropy we plotted the absolute value of the anisotropy. The absolute value was picked because it avoids the issue of sudden changes in sign which can happen when two eigenvalues cross each other due to the way the Haeberlen convention is defined. Figure \ref{fig:res_eth_aspirin_plot} shows the plot of $|\sigma_{aniso}|$ for all three protons, with their respective distances from the centre of the nearby ring marked with lines. It is easy to see that all three protons display strong new features at this range, starting usually a bit before the mark (since the mark labels the distance from the centre of the ring, the growing integration radius intersects the ring's electronic cloud a bit earlier). Predictably, the closest one displays the strongest effect.

%  the computed convergence radii for isotropic shieldings at all sites in ethanol and aspirin are shown at different tolerances. These can be considered to represent the effective size of the Lorentz sphere given a desired level of accuracy. The radii have been computed by finding the distance at which the error in the buildup function with respect to the final value for each site falls under the requested tolerance and never goes above again. The candlesticks represent the range between the minimum and maximum radius in a system. The values from Table \ref{tab:res_lattice} are plotted as well for reference. Within the estimated Lorentz sphere radii all sites seem to converge within at least 5 ppm of their final shielding values, while smaller fluctuations of a few ppm cease only on much larger scales, up to more than 7 \AA{}.

\begin{figure}[ht]
    \centering
    \includegraphics[width=0.5\textwidth]{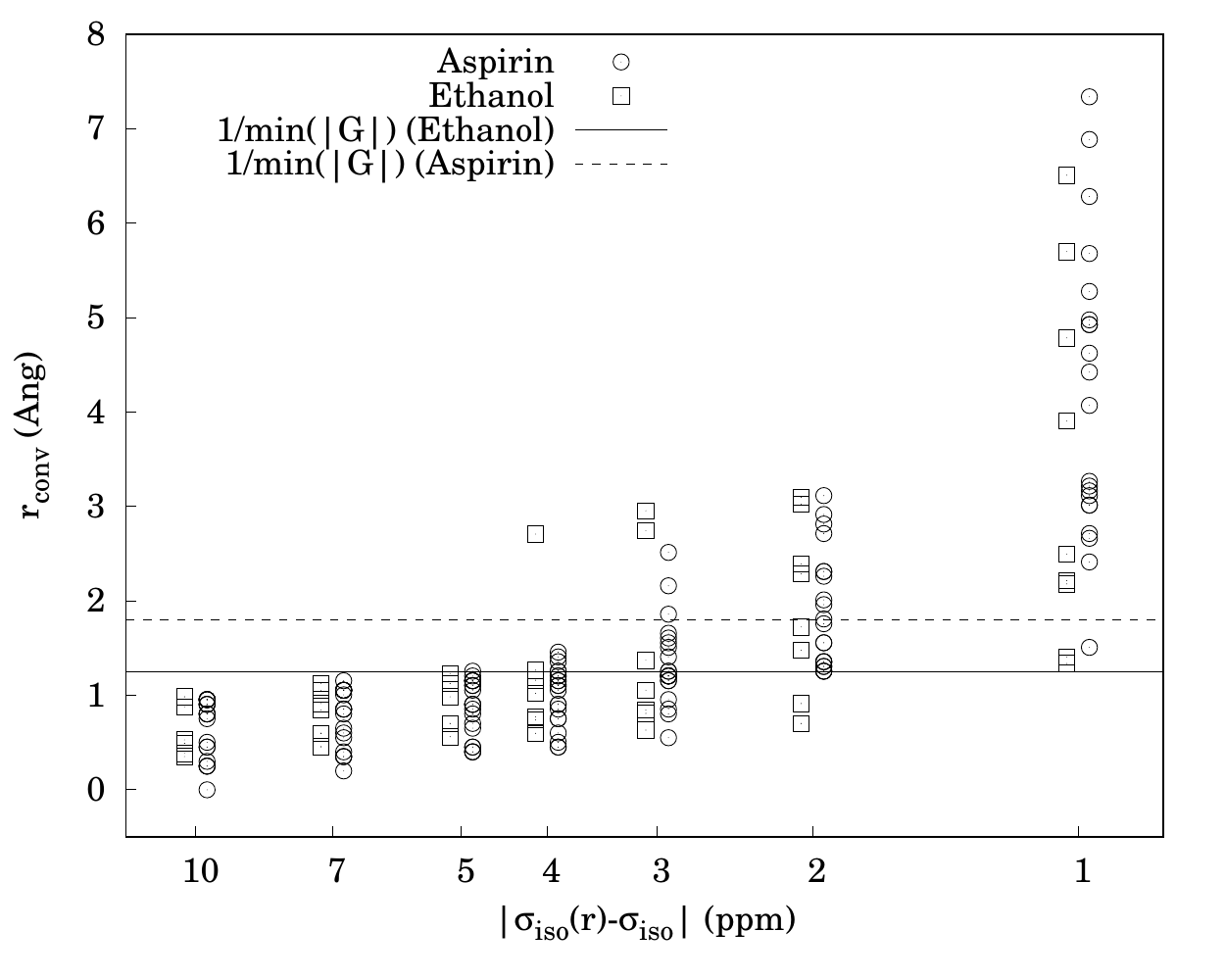}
    \caption{Convergence radii for all nuclear shieldings in ethanol and aspirin at different tolerances. The values of $1/\min(|G|)$ for both are plotted for reference. The limits appear to hold well for most nuclei for an accuracy down to 4 ppm, while radii up to more than 7 \AA{} are necessary to reach an accuracy of 1 ppm.}
    \label{fig:res_lorentzconv}
\end{figure}

Finally, Figure \ref{fig:res_lorentzconv} provides a clearer overview of the rate at which the shieldings converge. For all atoms in both ethanol and aspirin we can see how the size of the sphere necessary to achieve convergence grows as the tolerance is lowered. Save for a few outliers, the rough estimates based on $1/min(|G|)$ account for convergence up to a 4 ppm precision. The size tends then to grow much faster for smaller tolerances.

\section{\label{sec:concl}Conclusions}

An analytic method to compute a radial buildup function for magnetic shielding starting from the electronic current density calculated by electronic structure software in periodic systems has been provided. This method allows us to have a rough estimate of the scale of what is known as the Lorentz sphere for a given crystal, and to make more accurate estimates for specific sites. Observation of practical cases for two molecular crystals confirms this technique's potential for quantifying the effects of chemical groups and molecules based on their distance, thus giving us a deeper understanding of how crystal structure contributes to NMR spectra. Regardless of the specific examples used in this work, this technique is applicable to periodic systems in general, including inorganic framework materials. A rigorous definition of the size of the Lorentz sphere is proposed, and it is shown that this effectively corresponds to a rather short range for most chemical sites in the tested systems. This confirms some long held beliefs in the field of NMR spectroscopy that however have been historically hard to express in a quantitative way. 

\section{\label{sec:ackn}Thanks and acknowledgements}

Computing resources provided by STFC Scientific Computing Department's SCARF cluster. This work was conducted within the framework of the CCP for NMR crystallography, which is funded by the EPSRC grants EP/J010510/1 and EP/M022501/1.

\bibliographystyle{unsrt}
\bibliography{main}

\end{document}